\documentclass[a4paper,11pt]{article}
\pdfoutput=1 

\usepackage{jcappub} 
\usepackage[english]{babel}
\pdfoutput=1
\usepackage[utf8x]{inputenc}
\usepackage{epsfig,latexsym,cancel,amssymb,amsmath,xcolor}
\usepackage{hyperref}         
\usepackage{psfrag,rotating}
\usepackage{lscape}
\usepackage{enumerate}
\usepackage{amsfonts}

\renewcommand{\thetable}{$\arabic{table}$}

\newcommand{\ad}[1]{{\color{black}#1}}

\newcommand{\Mpcinv}{\,\mathrm{Mpc}^{-1}}
\newcommand{\gev}{\,\mathrm{GeV}}
\newcommand{\mev}{\,\mathrm{MeV}}
\newcommand{\cms}{\,\mathrm{cm^3 s^{-1}}}
\newcommand{\ms}{\,M_\odot}
\newcommand{\fpbh}{f_\mathrm{PBH}}

\newcommand{\amd}{a_\mathrm{dom}}

\newcommand{\arh}{a_\mathrm{rh}}

\newcommand{\mrh}{M_\mathrm{rh}}
\newcommand{\mmax}{M_\mathrm{max}}
\newcommand{\mmin}{M_\mathrm{min}}
\newcommand{\Hrh}{H_\mathrm{rh}}
\newcommand{\trh}{T_\mathrm{rh}}

\newcommand{\mphi}{m_\phi}
\newcommand{\mchi}{m_\chi}
\newcommand{\fdm}{f_\mathrm{DM}}
\newcommand{\fann}{f_\mathrm{ann}}
\newcommand{\fdec}{f_\mathrm{decay}}
\newcommand{\fligo}{f_\mathrm{LIGO}}
\newcommand{\sv}{\langle\sigma v\rangle}
\newcommand{\Hdom}{H_\mathrm{dom}}
\newcommand{\Nemd}{N_\mathrm{EMD}}
\newcommand{\sigmav}{\langle\sigma v\rangle}

\def\beq{\begin{equation}}
\def\eeq{\end{equation}}
\def\bea{\begin{eqnarray}}
\def\eea{\end{eqnarray}}

\def\ln#1{\mathrm{ln}\left(#1\right)}

\def\tev{\,\rm TeV}

\title{Solar Mass Primordial Black Holes in Moduli Dominated Universe}
    
\author[a,b]{Sukannya Bhattacharya,}
\author[c]{Anirban Das,}
\author[d]{and Koushik Dutta} 
\affiliation[a]{Theoretical Physics Division, Physical Research Laboratory, Navrangpura, Ahmedabad - 380009, India.}
\affiliation[b]{Centre for Strings, Gravitation and Cosmology, Department of Physics, Indian Institute of Technology Madras, Chennai 600036, India}
\affiliation[c]{SLAC National Accelerator Laboratory, 2575 Sand Hill Road, Menlo Park, California 94025, USA.}
\affiliation[d]{Indian Institute of Science Education and Research Kolkata, Mohanpur, WB 741246, India.}
\emailAdd{sukannya@physics.iitm.ac.in}
\emailAdd{anirband@slac.stanford.edu}
\emailAdd{koushik.physics@gmail.com}

\abstract{We explore the prospect of producing primordial black holes around the solar mass region during an early matter domination epoch. The early matter-dominated epoch can arise when a moduli field comes to dominate the energy density of the Universe prior to big bang nucleosynthesis. The absence of radiation pressure during a matter-dominated epoch enhances primordial black hole formation from the gravitational collapse of primordial density fluctuations. In particular, we find that primordial black holes are produced in the $0.1-10~M_{\odot}$ mass range with a favorable choice of parameters in the theory. However, they cannot explain all of the merger events detected by the LIGO/Virgo gravitational wave search. In such a case, primordial black holes form about $4\%$ of the total dark matter abundance, of which $95\%$ belongs to the LIGO/Virgo consistent mass range. The rest of the dark matter could be in the form of particles that are produced from the decay of the moduli field during reheating. }

\subheader{\hfill SLAC-PUB-17574}
\begin{document}

\maketitle


\section{Introduction}\label{intro}
One of the most compelling and exciting mysteries in contemporary cosmology is the nature of the dark matter (DM), which contributes about $\sim 26\%$ of the total energy budget of our Universe. Observational studies have revealed that DM is non-relativistic and interacts very \emph{weakly}, if at all, with the particles in the visible sector. So far, we have only been able to observe its gravitational influence on the visible sector of the Universe. Myriads of theories have been proposed to explain DM. Weakly interactive massive particle (WIMP), axion and axion-like particle, primordial black hole (PBH), dark macroscopic objects etc. are among the popular theories. However, we have not yet seen any compelling experimental or observational signature of any of these theories, and it still remains to be one of the most active fields of research.

Theories considering PBHs as DM candidate have gained significant interest recently after the observations of gravitational waves from binary black hole mergers by the LIGO/Virgo collaboration because of the possibility that the observed black holes of supersolar mass can be of primordial origin. PBHs were proposed as a DM candidate almost four decades ago\,\cite{Hawking:1971ei,Hawking:1974rv,Carr:1974nx,Carr:1975qj}. They can form due to gravitational collapse of large density fluctuations in the early Universe. Their mass can be calculated from the horizon mass of the Universe at the time of their formation. Depending on the formation epoch, their mass can span over many orders of magnitude. 

Unlike particle DM, the macroscopic size of PBHs demands for observational studies with qualitatively different nature. A variety of observational and experimental data now constrain a significant part of the PBH parameter space. Nevertheless, these constraints are expected to evolve in near future with advent of newer data. Due to Hawking radiation, PBHs evaporate on a timescale $t_{\rm ev}=5120\pi G^2M^3/(\hslash c^4)$, so that PBHs of mass lower than $M\simeq 5\times 10^{14}~{\rm g}$ or $2.5\times 10^{-19}\ms$ have completely evaporated by now\,\cite{Hawking:1974rv}. Slightly heavier PBHs have lower temperature, however, they have not completely evaporated yet and may radiate gamma-ray photons, neutrinos, gravitons and other massive particles at different stages of evaporation. Therefore, constraining the injection of photons and neutrinos in the (extra-)galactic medium using Voyager data, extra-galactic radiation background, SPI/INTEGRAL observations etc.~\cite{Churazov:2010wy,Siegert:2016ijv,Laha:2019ssq,Bays:2011si,Collaboration:2011jza,Agostini:2019yuq,Dasgupta:2019cae,Laha:2020ivk} light PBHs with $M \lesssim 10^{-17}\ms$ can be constrained. Abundance of such light PBHs is also constrained from CMB anisotropies (constrain PBHs with $M \geq 1.1\times 10^{13}$ g) and abundance of light elements (constrain PBHs with $M \simeq 2\times 10^{11} - 10^{12}$ g) at the time of nucleosynthesis due to the energy decomposition in the background by the evaporation products from the black holes~\cite{Acharya:2020jbv}. PBHs in the mass range $10^{-11}\ms<M<10^{-1}\ms$ are constrained by the gravitational lensing of light rays from distant stars by them. Observation of the stars in the M31 galaxy by the HSC telescope, the EROS and OGLE survey together now rule out the contribution of PBH towards total DM density above $1 - 10\%$ in this mass range\,\cite{Smyth:2019whb,Tisserand:2006zx,Niikura:2019kqi}. The caustic crossing event for the star Icarus or MACS J1149LS1 and the resultant strong lensing has been used to place constraint on compact objects in the range $10^{-5}\ms<M\lesssim 10^3\ms$\,\cite{Oguri:2017ock}. The GW detections by the LIGO/Virgo collaboration puts an upper bound on $\fpbh$ in the mass region $0.2\ms<M<300\ms$ assuming that the observed binary mergers events are PBH mergers in early or late Universe\,\cite{Ali-Haimoud:2017rtz,Bird:2016dcv,Sasaki:2016jop,Cholis:2016kqi,Clesse:2016vqa,Raccanelli:2016cud,Kovetz:2017rvv,Authors:2019qbw,Kavanagh:2018ggo,DeLuca:2020qqa,Wang:2016ana}. Finally, the radiation from the accreted gas around PBHs of mass $M \gtrsim 100\ms$ affects the spectrum and the anisotropies of the CMB\,\cite{carr1981pregalactic,Ricotti:2007au,Serpico:2020ehh}. See Ref.\cite{Carr:2020gox} for a recent review of the constraints on PBHs.

Everything considered, current observations have already constrained the PBH population in a significant part of its mass range meaning that it cannot form more than a few percent of total DM abundance in a large part of the parameter space. This leaves us with the possibility of a scenario where the dark sector of the Universe is comprised of multiple components. For comparison, if we look at the visible sector, we find both gravitationally collapsed objects, like stars, planets, black holes etc., and the diffused interstellar medium, cosmic dust, gas clouds which do not form macroscopic objects. Drawing this comparison, it would be natural to imagine a dark (matter) sector scenario where it consists of a combination of PBH and particle DM.

In this work, we consider an early matter dominated (EMD) epoch where the dominant energy contribution comes from moduli fields (for reviews, see \cite{Kane:2015jia, Allahverdi:2020bys}). In various string theory inspired models of inflation, moduli field is inactive during inflationary epoch at its metastable vacuum state. After the end of inflation, the Hubble parameter decreases, and owing to the `vacuum misalignment' of the moduli field, it starts to oscillate around its true minimum. During oscillation, the moduli field behaves like \emph{matter}. Such a scenario leads to a moduli dominated epoch after inflation for a period of time with a matter-like equation of state of the Universe before the big bang nucleosynthesis (BBN). We know from observations that the Universe was radiation-dominated during BBN. At the end of this epoch, the moduli field decays into other particles and \emph{reheats} the Universe, initiating the radiation-dominated epoch. A period of matter-domination has implications for the formation of PBH in the early Universe\,\cite{Khlopov:1980mg,Harada:2016mhb,Carr:2017edp,Ballesteros:2019hus,Ballesteros:2018wlw,Georg:2016yxa,Georg:2017mqk,Hidalgo:2017dfp}. 
	
The \emph{softening} of the equation of state implies a reduction in radiation pressure, and a subsequent enhancement in the probability of gravitational collapse of large density fluctuations into black holes. Therefore, the PBH formation rate increases significantly during such early matter domination relative to radiation domination. In this work, we investigate the possibility of forming $\sim 1-100\ms$ black holes during such EMD era, which could explain the binary black hole merger events detected by LIGO/Virgo. Most of the black holes detected by the gravitational wave searches have small angular momentum which is not expected if they are astrophysical in origin~\cite{Bird:2016dcv,Sasaki:2016jop}. This motivates us to look for alternative origin of these black holes such as production during an EMD epoch. We compute the mass function of the PBHs in this scenario and compare with the observational constraints. We find that it is possible to conceive of a scenario where PBHs are preferably produced in the mass range of $\sim 0.1-10\ms$ during the EMD epoch, and form about 95\% of all PBHs. This is achieved after saturating the BBN lower limit on the reheat temperature, i.e., when the EMD era ends just prior to BBN. However, as most of the LIGO/Virgo detections are above $10\ms$, this scenario can explain only a few of the observed merger events.

\ad{The structure of this paper is as follows. In section\,\ref{sec:moduli}, we introduce and discuss about the moduli field and its evolution. In section\,\ref{sec:pbh_production}, we review the PBH formation mechanism and the resulting mass functions in both radiation dominated (RD) and matter dominated (MD) Universe. We also discuss the primordial power spectrum as an input in calculating the PBH abundance and the current observational constraints on such theoretical predictions. Then in section\,\ref{sec:particle_production}, we outline the possible ways to produce particle DM from the decay of the moduli field and the mechanisms to form a relic abundance of these particles. Finally in section\,\ref{sec:analysis}, we take all constraints into account and analyze the viable range of various parameters in this framework, and conclude in section\,\ref{sec:conclusions}.}

\section{Moduli domination and reheating}\label{sec:moduli}
\ad{In this section, we shall review the nonstandard thermal history of the early Universe due to dynamical effects of a moduli field. Moduli is a scalar field $\phi$ with its post-inflationary mass $\mphi$ with potential  $V (\phi)=m_\phi^2 \phi^2/2$. To begin with, at the end of inflation the moduli field is frozen at its initial value $\phi_0$. It starts to move when the Hubble parameter approximately equals its mass, $H\simeq m_\phi$. Afterwards, it continues oscillating about the minimum of its potential, and the energy density carried by the field redshifts as $1/a^{3}$ where $a$ is the scale factor of the Universe. Soon after, the energy density of the moduli starts to dominate the energy density of the Universe, marking the onset of a matter dominated Universe. 

Finally, the moduli decays into visible and dark sector particles to produce a thermal bath of temperature that is suitable for big bang nucleosynthesis (BBN). Typically, the decay width of a moduli field is given by~\cite{Cicoli:2016olq}
	\begin{equation}\label{eq;decay_width}
		\Gamma \simeq \dfrac{\mphi^3}{YM_\mathrm{Pl} ^2}\,.
	\end{equation}
	where $M_\mathrm{Pl} = 2.44\times 10^{18}$ GeV is the reduced Planck mass, and $Y$ is a model dependent parameter which depends on the origin of the moduli field in a fundamental theory. In this work, we will consider $Y = 16\pi$, and our results remain qualitatively unchanged for other plausible values of $Y$~\cite{Cicoli:2010ha,Conlon:2007gk,Cicoli:2012aq,Higaki:2012ar}. While oscillating, the energy density of $\phi$ behaves like matter. It may come to dominate the total energy density of the Universe when $\Hdom \simeq m_\phi(\phi_0/M_\mathrm{Pl})^4$. After the moduli decays into other particles when $\Hrh\simeq \Gamma$, the Universe reheats, and becomes radiation-dominated again. 
	
The epoch of BBN puts a lower limit on the reheat temperature $\trh\gtrsim 4.3\mev$\,\cite{deSalas:2015glj}. Therefore, the moduli needs to decay before this epoch not to modify the observed abundance of primordial elements. The reheat temperature $\trh$ is then given by
	\begin{equation}\label{eq:trh1}
		\trh = \left(\dfrac{90}{\pi^2 g_*(\trh)}\right)^{1/4} \sqrt{\Gamma M_\mathrm{Pl}} = 2.75\mev\,\left(\frac{10.66}{g_*(\trh)}\right)^{1/4} \left(\frac{m_\phi}{100 \text{TeV}}\right)^{3/2} \,,
	\end{equation}
	where $g_*(T)$ is the number of relativistic degrees of freedom at photon temperature $T$, and we use $g_*(T=4.3~ \text{MeV}) = 10.68$. The BBN bound on $\trh$ translates into a lower limit for $\Gamma$, which, from Eq.~\eqref{eq:trh1}, gives $\Gamma \gtrsim 8\times 10^{-24}\gev$, and using the decay width from Eq.~\eqref{eq;decay_width}, it translates into $\mphi \gtrsim 135$ TeV. 
	
The epoch of EMD spans from $\Hdom$ until $\Hrh$ with $\Nemd$ number of e-foldings,
	\begin{equation}
		\Nemd = \ln{\frac{\arh}{a_\mathrm{dom}}} \,. 
	\end{equation}
In the moduli sector, there are only two scales, namely $m_\phi$ and $\phi_0$, that determine the reheat temperature and the duration of the EMD epoch. In this work, we shall fix $\phi_0=0.05 M_\mathrm{Pl}$ for concreteness~\cite{Cicoli:2016olq,Allahverdi:2018iod,Gallego:2020vbe}. Our final results for the total PBH abundance are only mildly sensitive to this choice as discussed in later sections. In the following sections, we shall choose our benchmark value of $m_\phi$ in such a way that it saturates the BBN bound, and provides the maximum duration for the EMD epoch. 
}

\section{Primordial black hole production}\label{sec:pbh_production}

The large density fluctuations required for PBH formation can originate from large primordial scalar fluctuations produced during inflation. In the linear order, the density contrast $\delta = \frac{\Delta \rho}{\rho}$ and inflationary curvature perturbation $\zeta$ are related as~\cite{Kalaja:2019uju}
\begin{equation}
	\delta ({\bf x},t)=\frac{2(1+w)}{5+3w}\bigg( \frac{1}{aH}\bigg)^2\bigtriangledown ^2\zeta ({\bf x},t),
	\label{delta1}
\end{equation}
where $w$ is the equation of state of the background and $a$ is the scale factor.

If the density fluctuations are larger than a critical density $\delta _c$, then they collapse gravitationally to form a black hole of mass $M$, which is typically of the order of the horizon mass $M_H$ at the time of the reentry of a mode $k=aH$, given as
\begin{equation}
	M=\gamma M_H= \gamma \frac{4 \pi M_\mathrm{Pl}^2}{H}.\label{mass1}
\end{equation}
The factor $\gamma$ depends on the exact details of the collapse process and is typically $\mathcal{O}(1)$. The critical density contrast $\delta_c$ depends on the background equation of state $w$. Early literatures (see Ref.~\cite{Carr:1975qj}) have considered $\delta_c=w$ with a discussion of uncertainties of order unity in the numerical prefactor of $w$ and subsequently, $\delta_c (w)$ has been studied numerically in~\cite{Niemeyer:1999ak,Musco:2004ak}. In this paper, we will use the following expression for $\delta_c (w)$, which was analytically developed and numerically fitted in Ref.~\cite{Harada:2013epa},
\begin{equation}
	\delta_c = \dfrac{3(1+w)}{5+3w}\sin^2\left(\dfrac{\pi\sqrt{w}}{1+3w}\right)\,.\label{deltac1}
\end{equation}
Therefore, in a radiation dominated (RD) epoch, $\delta_c (w=1/3) \simeq 0.414$.

For a Gaussian distribution of the fluctuations $\delta$, the probability of the gravitational collapse to a PBH of mass $M$ is given by the mass fraction (using Press-Schechter formalism)
\begin{equation}\label{eq:beta}
	\beta(M) = \int_{\delta_c}^{\infty} d\delta ~P(\delta) = \frac{2}{\sqrt{2\pi}\sigma(M)} \int^\infty_{\delta_c} d\delta \exp\left(-\frac{\delta^2}{\sigma(M)^2}\right) = \mathrm{erfc}\left(\frac{\delta_c}{\sqrt{2}\sigma(M)}\right)\,.
\end{equation}
Here $\sigma ^2(M)$ is the variance of the density fluctuation at a mass scale $M$ and can be written in terms of the primordial curvature power spectrum $P_{\zeta}(k)$ as
\begin{equation}
	\sigma^2(M)=\frac{4(1+w)^2}{(5+3w)^2}\int \frac{dk}{k}(kR)^4 W^2(k,R)P_{\zeta}(k),\label{sigmaM1}
\end{equation}
where $P_{\zeta}(k)$ is the primordial power spectrum of curvature perturbations.
Choosing a Gaussian window function $W(k,R)$ smoothens the perturbations on the comoving scale $R$ at formation. However, we use the following approximate relation
\begin{equation}\label{sigmaM2}
	\sigma (M) \simeq \frac{2(1+w)}{(5+3w)}\sqrt{P_{\zeta}(k)}
\end{equation}
where the relation between $M$ and wavenumber $k$ can be obtained from Eq.~\eqref{mass1} given a particular background evolution. 

However, the expression for $\delta _c (w)$ in Eq.~\eqref{deltac1} fails for exact matter domination with $w=0$. Ref.~\cite{Harada:2016mhb} analytically formulated the dynamics of PBH formation in matter dominated epoch using Zel'dovich approximation and found the numerical fit of the mass fraction to be $\beta (M) \simeq 0.056~ \sigma(M)^5$, which we will use for the MD case here. 

Scalar fluctuations of large amplitudes at small scales during inflation can lead to copious amount of PBH formation, which is evident from the dependence of $\sigma (M)$ on $P_{\zeta}(k)$. In single field slow roll models of inflation, such large scalar fluctuations can arise from certain features in the inflationary potential, such as inflection points~\cite{Garcia-Bellido:2017mdw,Bhaumik:2019tvl}, tiny bump or dip~\cite{Mishra:2019pzq} etc. In multifield models of inflation, features in $P_{\zeta}(k)$ can be induced by kinetic coupling with a secondary field~\cite{Braglia:2020eai}, or non-geodesic motion in field space induced by a large turn of the inflaton~\cite{Fumagalli:2020adf} etc. PBH formation in curvaton inflation scenarios have also been explored~\cite{Kohri:2012yw,Ando:2018nge} whereas PBH from warm inflation is addressed in Ref.~\cite{Arya:2019wck}. Although inflation offers the simplest mechanism to produce PBHs via small-scale enhancement of $P_{\zeta}(k)$, there are other mechanisms of PBH formation, such as bubble nucleation~\cite{Ashoorioon:2020hln} collapse of domain walls~\cite{Deng:2016vzb,Liu:2019lul,Deng:2017uwc}, Q-balls~\cite{Cotner:2016cvr,Cotner:2017tir} etc. PBH formation mechanism also depends strongly on the background epoch at the time of collapse. Other than the pioneering works for PBH formation in early matter domination in Ref.~\cite{Khlopov:1980mg,Polnarev:1985,Harada:2016mhb}, PBH formation from slow reheating~\cite{Carr:2018nkm} ($w$ changing slowly from 0 to 1/3), stiff dominated epochs~\cite{Bhattacharya:2019bvk} ($1/3 <w<1$) etc. have also been explored.

The fraction of the background energy density that collapses into forming a PBH is $\frac{\rho_{\rm {PBH}}(M)}{\rho}\vert _{i} = \gamma \beta (M) $ where the subscript $i$ defines time of formation of PBH of mass $M$, and $\rho$ is the total energy density of the Universe at the time of formation\footnote{This fraction is simply $\beta (M)$ for matter domination, since here, the separation between time of horizon entry and time of collapse is accounted for in the description of mass, as explained later.}.
The fraction of DM in the form of PBH, i.e., PBH abundance, is defined through the PBH mass function $\psi(M)$ as
\begin{equation}\label{eq:psi}
	\psi (M)=\frac{1}{M}\frac{\Omega_{\rm PBH}(M)}{\Omega _{\rm DM}}\bigg \vert _0\,.
\end{equation}
To compare with observations, we define $\psi (M)$ at the present epoch (subscript `0'), which also depends on cosmological evolution from PBH formation until today.  The total abundance of PBH is therefore
\begin{equation}
	f_{\rm PBH}=\int \psi (M) dM. \label{fpbh00}
\end{equation}

\subsection{Production in standard radiation domination}
In a standard cosmological history with no additional epoch of MD, PBHs are produced during radiation domination after the end of inflation. Thus, the mass function for PBH of mass $M$ is
\begin{equation}
	\begin{array}{rcl}
	\psi(M) &=& \dfrac{1}{M} \dfrac{\rho_{\rm PBH}(M)}{\rho_{\rm DM}}\bigg\vert _{0} = \dfrac{1}{M} \dfrac{\rho_{\rm PBH}(M)}{\rho_{\rm rad}}\bigg\vert _{\rm eq}\dfrac{\Omega_{\rm m}^0h^2}{\Omega_{\rm DM}^0h^2} = \dfrac{\Omega_{\rm m}^0h^2}{\Omega_{\rm DM}^0h^2} \dfrac{\gamma \beta _{\rm RD}(M)}{M}\dfrac{a_{\rm eq}}{a_{i}}\,.
	\end{array}
\end{equation}
Here, $\beta _{\rm RD}(M)$ is calculated using Eq.~\eqref{eq:beta} with $\delta _c=0.414$ and $\sigma (M)$ from Eq.~\eqref{sigmaM2} as inputs. 
During RD, $\frac{a_{\rm eq}}{a_{i}}=\big(\frac{H_i}{H_{\rm eq}}\big)^{1/2}=\big(\frac{M_{\rm eq}}{M}\big)^{1/2}$ so that
\begin{equation}
\begin{array}{rcl}
	\psi(M) &=& \dfrac{\Omega_{\rm m}^0h^2}{\Omega_{\rm DM}^0h^2}\bigg(\dfrac{M_{\rm eq}}{M_{\odot}}\bigg)^{1/2}\bigg(\dfrac{M_{\odot}}{M}\bigg)^{1/2} \dfrac{\gamma \beta _{\rm RD}(M)}{M} \simeq 5\times 10^8~\dfrac{\beta _{\rm RD}(M)}{M}\bigg(\dfrac{M_{\odot}}{M}\bigg)^{1/2},
	\end{array}
	\label{psiRD1}
\end{equation}
where, for the last equality we have used Planck 2018 values $\Omega_{\rm m}^0=0.315$, $\Omega_{\rm DM}^0h^2=0.12$ and $h=0.67$\,\cite{Aghanim:2018eyx}. We consider $\gamma = 0.4$ in our analysis.

\subsection{Production in presence of moduli domination}
If there is an additional post-inflationary epoch dominated by moduli fields, then PBHs can be formed before, during, and after moduli domination, where the background evolutions follow RD, MD and RD respectively. In this section, we derive the relation between $\psi(M)$ and $\beta(M)$ for each of these epochs in presence of moduli domination.

For the modes that entered during EMD, the mass function is
\begin{equation}
\begin{array}{rcl}
	\psi(M) &=& \dfrac{1}{M} \dfrac{\rho_{\rm PBH}(M)}{\rho_{\rm DM}}\bigg\vert _{0} = \dfrac{1}{M} \dfrac{\rho_{\rm PBH}(M)}{\rho_{\rm rad}}\bigg\vert _{\rm eq}\dfrac{\Omega_{\rm m}^0h^2}{\Omega_{\rm DM}^0h^2} = \dfrac{\Omega_{\rm m}^0h^2}{\Omega_{\rm DM}^0h^2} \dfrac{\beta _{\rm MD}(M)}{M}\dfrac{a_{\rm eq}}{a_{\rm rh}}\,.
	\end{array}
\end{equation}
Now, $a\sim 1/\sqrt{H}$ during RD from reheating (`rh') until equality, $H_\mathrm{rh}\simeq \Gamma$, and writing $H_\mathrm{eq}$ in terms of $M_\mathrm{Pl}$ gives
\begin{equation}\label{eq:emd_mf}
	\psi(M) \simeq 3.66\times 10^{27}\left(\dfrac{\Gamma}{M_\mathrm{Pl}}\right)^{1/2} \dfrac{\beta _{\rm MD}(M)}{M}\,.
\end{equation}

The difference between PBH formation in MD and RD epoch lies in the different collapse dynamics in these two epochs. The absence of pressure can lead to non-spherical effects, which is discussed in detail in Ref.~\cite{Harada:2016mhb}, where for perturbations of order $\sigma \lesssim 0.01$, the mass fraction was found to be
\begin{equation}\label{eq:beta_emd}
\beta _{\rm MD}(M)\simeq 0.056\sigma(M)^5.
\end{equation}
During MD, density perturbations grow linearly so that $\delta \simeq \sqrt{\langle \sigma^2 \rangle} \sim a$. It becomes nonlinear when $\delta \simeq \mathcal{O}(1)$. If $\sigma$ is defined in the linear regime at the time of horizon entry of the modes, then the scale factor $a_m$ at the time $t_m$ of maximum expansion is given by $\sigma (a_m/a_H)=1$, where $a_H$ is the scale factor at the time $t_H$ of horizon entry~\cite{Carr:2017edp,Harada:2017fjm,Nakama:2018utx,Khlopov:1980mg,Polnarev:1985}. The time of collapse $t_c$ is very close to $t_m$ and therefore, the scale factor at $t_c$ is $a_c\simeq a_m$. Thus, $t_c/t_H=(a_c/a_H)^{3/2}=\sigma ^{-3/2}$. The Hubble parameters at horizon entry ($H_{\rm hor}$) and at the time of collapse ($H_c$) are therefore related as $H_{\rm hor}/H_c=\sigma ^{-3/2}$. So, the PBH that is formed from the mode that enters the horizon at $t_H$ has a mass
\begin{equation}\label{genmassEMD}
 M=\frac{4 \pi M_\mathrm{Pl} ^2}{H_{\rm hor}}=\frac{4 \pi M_\mathrm{Pl} ^2}{H_c}\sigma ^{3/2}~.
 \end{equation}
In the EMD epoch, the PBH mass formed as a result of collapse (when $\sigma$ becomes nonlinear) at time $t_c$ can be estimated using the last equality in Eq.~\eqref{genmassEMD}. In comparison, note that the growth of perturbations is logarithmic in a RD epoch, and therefore, the mass of PBHs formed in RD can be estimated by Eq.~\eqref{mass1}. 

The PBH mass function in EMD given by Eq.~\eqref{eq:emd_mf} is limited within two mass scales, $\mmax$ and $\mmin$ corresponding to the largest and the smallest scales, respectively, that became nonlinear during EMD. The largest mass scale $\mmax$ corresponds to the mode $H_{\rm max}^{-1}$ that entered the horizon at some point before reheating and collapsed at the time of reheating. Therefore, following the arguments in the previous paragraph, $\mmax$ is given by~\cite{Carr:2017edp}
\begin{equation}
	M_{\rm max} = \dfrac{4 \pi M_\mathrm{Pl}^2}{H_\mathrm{max}} = \frac{4 \pi M_\mathrm{Pl}^2}{H_{\rm rh}}\sigma_{\rm max}^{3/2} =\frac{4 \pi M_\mathrm{Pl}^2}{\Gamma}\sigma_{\rm max}^{3/2} = \mrh\sigma_{\rm max}^{3/2}\,,\label{eq:mmax}
\end{equation}
where we have used $H_{\rm rh} \simeq \Gamma$, and $\sigma_{\rm max}$ will be calculated from the primordial power spectrum as discussed later. 

The lower mass scale corresponds to the mode that became nonlinear at the begining of the EMD epoch. The moduli field starts to dominate the energy density at $H_{\rm dom}= (\phi_0/ M_\mathrm{Pl})^4 m_{\phi}$. Therefore, number of e-foldings of moduli domination is given by
\begin{equation}\label{Nemdex}
	\Nemd=\log\left(\frac{a_{\rm rh}}{a_{\rm dom}}\right)=\frac{1}{6}\log\left( \frac{H_{\rm dom}}{H_{\rm rh}} \right)=\frac{1}{6}\log\left(\frac{16 \pi \phi_0^4}{m_{\phi}^2 M_\mathrm{Pl}^2} \right)\,.
\end{equation}
Hence,
\begin{equation}
	M_{\rm min} = M_\mathrm{max} \left(\frac{a_{\rm dom}}{a_{\rm rh}}\right)^{3/2} = \mmax\left( \frac{\mphi M_\mathrm{Pl}}{ 4 \sqrt{\pi} \phi_0^2}\right)^{1/2} \,.
\end{equation}

Now, the modes that enter before and after the EMD epoch will both encounter a RD background, however, the respective mass functions will be different due to the non-trivial cosmological evolution. For modes that entered before the onset of the EMD epoch at $\amd$,
\begin{equation}
\begin{array}{rcl}
\psi (M)=&\dfrac{1}{M}\dfrac{\rho _{\rm PBH}}{\rho_{\rm rad}}\bigg \vert _i \dfrac{a_{\rm dom}}{a_{\rm rh}}\dfrac{a_{\rm eq}}{a_{\rm rh}}\dfrac{\Omega_{\rm m}^0h^2}{\Omega_{\rm DM}^0h^2} =& 5\times 10^8 \left(\dfrac{\ms}{M}\right)^{1/2} \left(\dfrac{\amd}{\arh}\right) \dfrac{\beta _{\rm RD}(M)}{M}, 
\end{array}
\end{equation}
where $(\amd/\arh)$ can be calculated from Eq.~\eqref{Nemdex}.
For modes that entered after $\arh$, the mass function is same as that in a standard RD epoch and is gven by Eq.~\eqref{psiRD1}.\footnote{The modes $H_{\rm max}< k/a < H_{\rm rh}$ have growth in both MD and RD epochs, however, their collapse always takes place in the RD epoch. This leads to the requirement of a separate analysis to derive $\psi (M)$ for these transient modes. However, since RD collapse leads to a smaller $\beta (M)$ than in MD, therefore, we consider their contribution to be negligible to the total abundance and proceed with a sharp drop in $\psi (M)$ between EMD and RD epochs.}:
 Therefore, the complete PBH mass function in all epochs reads as follows:
\begin{equation}
		\psi(M) =
	\begin{cases}
		2.6\times 10^8\left(\dfrac{\ms}{M}\right)^{1/2} \left(\dfrac{m_\phi M_\mathrm{Pl}}{\phi_0^2}\right)^{1/3} \dfrac{\beta _{\rm RD}(M)}{M}\,,\quad\,M<M_{\text{min}}\,,\\
		5.2\times 10^{26}\left(\dfrac{m_\phi}{M_\mathrm{Pl}}\right)^{3/2} \dfrac{\beta _{\rm MD}(M)}{M}\,,\quad M_{\text{min}}\leq M \leq M_{\text{max}} \,,\\
		5\times 10^8\left(\dfrac{\ms}{M}\right)^{1/2} \dfrac{\beta _{\rm RD}(M)}{M}\,,\quad\,M> M_{\text{max}}\,.
	\end{cases}
	\label{psiMall}
\end{equation}
We shall use this mass function to calculate the PBH abundance in section~\ref{sec:analysis}.


\subsection{Primordial power spectra}\label{sec:PPS}
The requirement of large primordial fluctuations for the production of abundant PBHs can be attained with several small-scale features in the inflationary paradigm. For single-field slow roll models of inflation, large amplitude of scalar fluctuations translates to ultra slow-roll of the inflaton, $\epsilon \ll \epsilon _{\rm CMB}<1$, where $\epsilon _{\rm CMB}$ is the typical value at the CMB pivot scale, due to the relation $P_{\zeta}(k)\propto \frac{1}{\epsilon}$. In models with an inflection point~\cite{Garcia-Bellido:2017mdw,Ballesteros:2017fsr,Ballesteros:2020qam,Bhaumik:2019tvl,Germani:2017bcs} at small scales, $\epsilon$ attains a tiny value at the point of inflection, offering a large amplitude of scalar perturbations. In models with a small bump (or dip) in the potential~\cite{Mishra:2019pzq}, the inflaton has negligible velocity at the top (or bottom) of the feature and hence leads to large $P_{\zeta}(k)$. For multifield models of inflation, the coupling of the inflaton with the secondary field can induce large $P_{\zeta}(k)$. As an example, for hybrid inflation models, this is achieved in the mild waterfall phase~\cite{Clesse:2015wea}. In some other cases, enhancement in $P_{\zeta}(k)$ results from noncanonical coupling~\cite{Braglia:2020eai,Braglia:2020taf} of the inflaton and the secondary field (leading to a second phase of inflation) or from large turning rate in the field space~\cite{Fumagalli:2020adf,Palma:2020ejf} inducing instabilities in the isocurvature fluctuations that can be transferred to curvature fluctuations. Instabilities arising in certain scenarios of effective field theories of inflation can also lead to large scalar power spectrum~\cite{Ashoorioon:2019xqc}.

\begin{figure}[t]
	\begin{center}
		\includegraphics[width=0.48\textwidth]{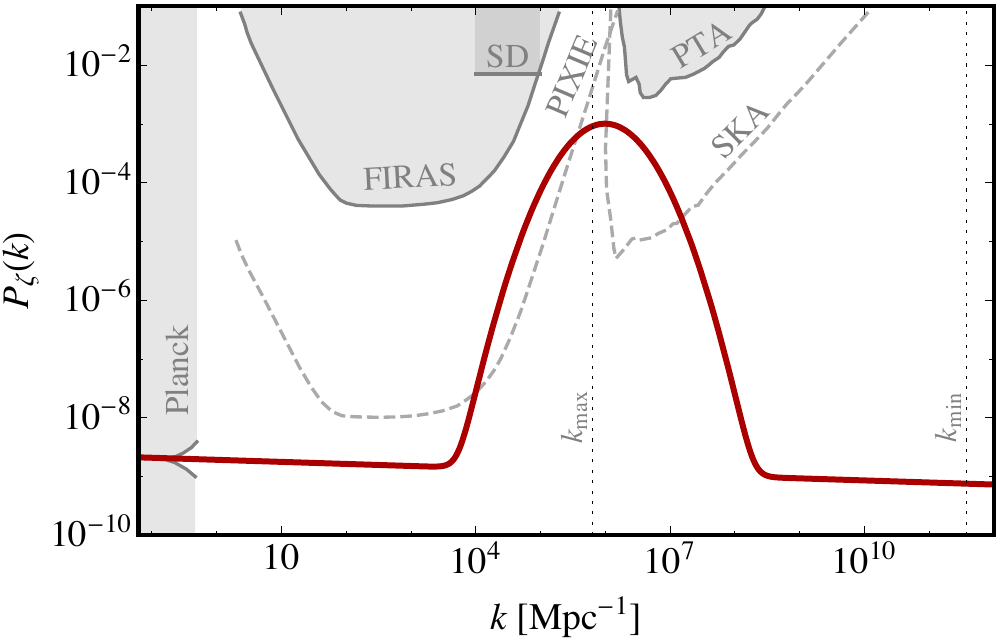}
		\hfill
		\includegraphics[width=0.48\textwidth]{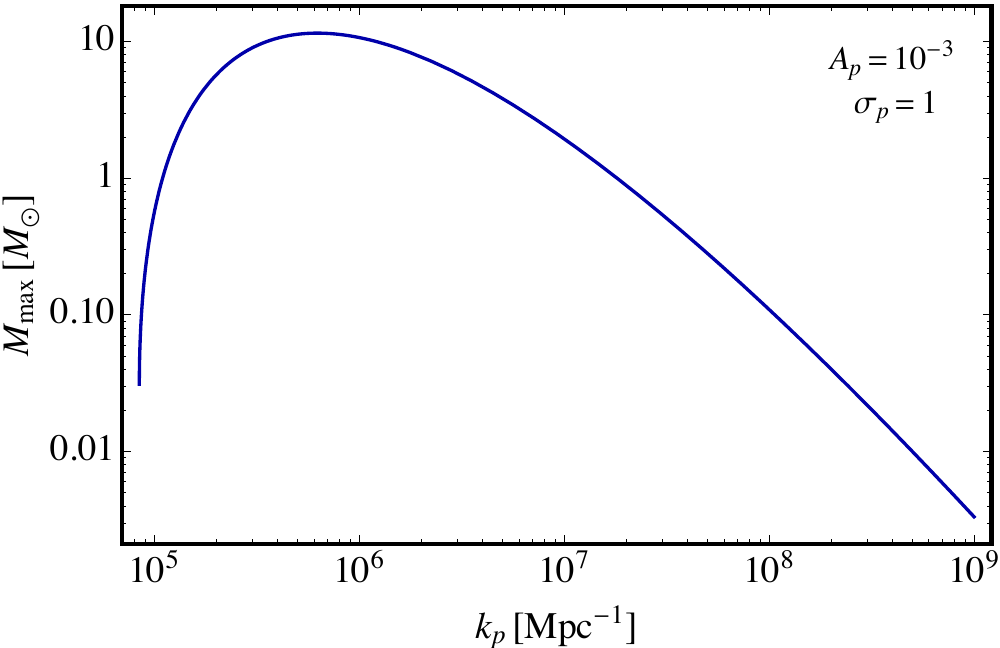}
		\caption{(Left) The primordial power spectrum $P_\zeta(k)$ for $A_p=10^{-3}, \sigma_p=1,$ and $k_p=10^6\Mpcinv$. The observational bounds on $P_\zeta(k)$ from present (future) experiments are shown with grey shaded (dashed lines) region.		
The vertical dashed lines correspond to the horizon entry scales for the minimum and the maximum masses formed during the EMD epoch. 
		 (Right) The variation of $\mmax$ with the pivot scale $k_p$ for the same $A_p$ and $\sigma_p$ as in the left panel.}\label{fig:primordial_spectrum}
	\end{center}
\end{figure}

In many such models of inflation, resulting $P_{\zeta}(k)$ can be well-approximated in the form of a Gaussian (in some cases, with an exponential cut-off~\cite{Vaskonen:2020lbd}). Therefore, for simplicity, we will work with the following $P_{\zeta}(k)$ which has a power enhancement at small scales due to a Gaussian profile centred at $k_p$ with an amplitude $A_p$, and width $\sigma_p$.
\begin{equation}
P_{\zeta}(k)=A_s\bigg(\frac{k}{k_*}\bigg)^{n_s-1} + A_p \exp\bigg[-\frac{(N_k-N_p)^2}{2\sigma _p^2}\bigg], \label{pow3}
\end{equation}
where $A_s=2.1\times 10^{-9}, n_s=0.965$, $k_*=0.05\Mpcinv$ for the CMB modes according to the Planck measurements  \,\cite{Aghanim:2018eyx}, and $N_k=\log (a(k)/a_{\rm end})$ is the number of e-folds before the end of inflation when the mode $k$ exits the horizon, so that $N_p=\log (a(k_p)/a_{\rm end})$. In figure\,\ref{fig:primordial_spectrum}, we show the primordial power spectra for $A_p=10^{-3}$,  $k_p=10^6\Mpcinv$ and $\sigma_p=1$. 

The parameter space of the primordial power spectrum is constrained by several experiments in various regimes of $k$. At large scales $k\lesssim 0.2\Mpcinv$, Planck data restricts the amplitude to $A_s=2.1\times 10^{-9}$ to a high precision. This is shown as the grey-shaded column at the extreme left of figure\,\ref{fig:primordial_spectrum}. Data from the FIRAS experiment looking for spectral distortion in the cosmic microwave background radiation and Pulsar Timing Array (PTA) data constrain the power spectrum amplitude in a large region of the $k$- space $k\simeq 10^2-10^4\Mpcinv$ (grey-shaded region at the top of figure\,\ref{fig:primordial_spectrum})\,\cite{Mather:1993ij,Moore_2014,Byrnes:2018txb}. Future mission PIXIE \cite{2011JCAP...07..025K}, which will carry out more precise measurement of the CMB spectral distortion, is expected to improve this bound by about four orders of magnitude. Similarly, future radio survey by SKA \cite{Moore_2014} will improve the current PTA constraint by about two orders of magnitude. 

Our choice of $k_p=10^6\Mpcinv$ is driven by our aim to form abundant PBHs with mass in the LIGO/Virgo region. The choices of $A_p$ and $\sigma _p$ are motivated by observational constraints on $P_{\zeta}(k)$, as well as by constraints on the PBH abundance by various observations, which will be discussed in detail in section~\ref{sec:analysis}. It is evident from figure~\ref{fig:primordial_spectrum} that present FIRAS and PTA bounds constrain the rise and fall of $P_{\zeta}(k)$ for our choice of $k_p$, i.e., it constrains $\sigma _p$. The choice of $\sigma _p =1$ along with the observational constraints on PBH abundance for the extended power spectrum in the left panel of figure~\ref{fig:primordial_spectrum} allows us to attain only $A_p \lesssim 10^{-3}$.\footnote{FIRAS data along with the condition for steepest growth of a power spectrum~\cite{Byrnes:2018txb} puts a lower limit on $\sigma _p$ for a single field inflation model. However, since we are ignorant to the exact details of the inflation model, this limit is not implemented strictly in our analysis.}

The wavenumbers $k_{\rm min}=k_{\rm rh}\exp (\Nemd/2)$ and $k_{\rm max}=k_{\rm rh}\sigma _{\rm max}^{-1/2}$ corresponding to the horizon-entry scales of $\mmax$ and $\mmin$ are also shown (for the above mentioned choice of parameters) in the left panel of figure~\ref{fig:primordial_spectrum}. To evaluate $k_{\rm min}$, $\Nemd$ can be determined using Eq.~\eqref{Nemdex}, where $\mphi = 135$ TeV and $\phi _0$ is kept fixed at $0.05 M_\mathrm{Pl}$. The position of $k_{\rm max}$ implies that $\mmax$ is very close to the peak of $P_{\zeta}(k)$, which hints at reaching large abundance for the intended $\mmax$ in the LIGO/Virgo mass range. As we see from the plot, the minimum PBH mass $\mmin$ corresponds to the mode with low amplitude $P_{\zeta}(k_{\text{min}})$ which depends on $\sigma _p$. However,  we note that $P_{\zeta}(k_{\rm min})$ can be larger for $\sigma _p >1$ (wider Gaussian profile), although, as it will be discussed in section~\ref{sec:analysis}, the total $f_{\rm PBH}$ depends very weakly on $\sigma _p$ when several observational bounds are considered.

It is interesting to note that, since the main contribution to the total PBH abundance comes from the mass function in the EMD epoch (middle expression in Eq.~\eqref{psiMall}), therefore, our analyis is mainly sensitive to the patch of $P_{\zeta}(k)$ between $k_{\rm max}$ and $k_{\rm min}$. For LIGO/Virgo mass PBHs (i.e. optimally large $\mmax$ here), this patch starts close to the Gaussian peak as can be seen in figure\,\ref{fig:primordial_spectrum}.   Looking at the maximum attainable mass in the EMD epoch (Eq.~\eqref{eq:mmax}), we can see that the quantity $\sigma _{\max}$ plays a crucial role. We consider $\trh$ to coincide with $T_{\rm BBN}=4.3$ MeV to have the maximum value for $M_{\rm rh}$, but the maximization of $\sigma _{\max}$ is also intended. From $P_{\zeta}(k)$ in Eq.~\eqref{pow3}, $\sigma _{\rm max}$ given by $\sigma _{\max}^2 \simeq\big(2/5\big)^2P_{\zeta}(k_{\rm max})$, which leads to
\begin{equation}
\log \sigma _{\rm max}=2\log\left(\frac{k_{\rm rh}}{k_p}\right)-8\sigma_p^2 \pm 2\sqrt{16\sigma_p^4-8\sigma_p^2\log\left(\frac{k_{\rm rh}}{k_p}\right)+2\sigma_p^2\log\left(\frac{4A_p}{25}\right)},
\end{equation}
and we use it in our calculations. We note that the $\sigma _{\rm max}$ depends on $k_p$, $A_p$, $\sigma _p$ and $\Gamma$ (through $k_{\rm rh}$) in a complicated way. In the right panel of figure~\ref{fig:primordial_spectrum}, the corresponding dependence of $\mmax$ on $k_p$ is shown, which reaffirms our choice of $k_p \simeq10^6\Mpcinv$ to reach $\mmax$ in the LIGO/Virgo mass range. We find $\sigma _{\rm max}\simeq 0.012$ for the benchmark values of $A_p$, $\sigma _p$ and $\Gamma (\trh=4.3 {~\rm MeV})$ as mentioned above. A better understanding for the choices of parameters in this section (and in figure\,\ref{fig:primordial_spectrum}) can be done in section \ref{sec:analysis}.

\begin{figure}
	\begin{center}
		\includegraphics[width=0.7\textwidth]{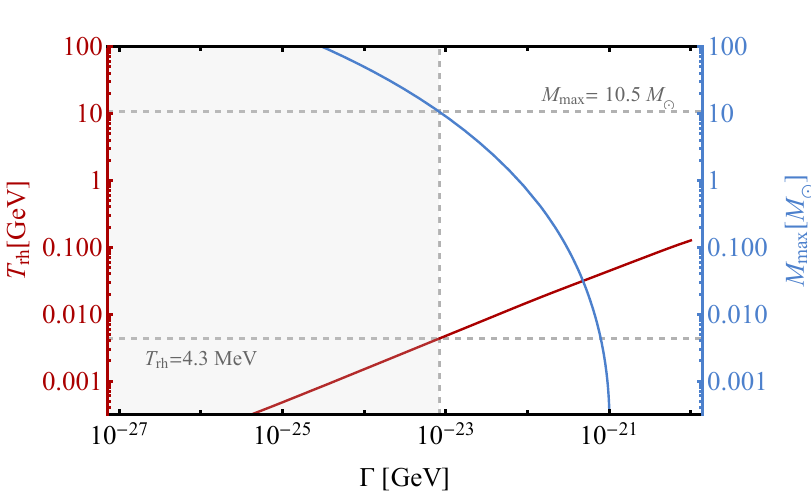}\label{mmax_plot}
		\caption{The variation of $\trh$ (red, left y-axis) and $\mmax$ (blue, right y-axis) as functions of the moduli decay width $\Gamma$ for $\sigma_{\text{max}} \simeq 0.012$. Big bang nucleosynthesis puts a lower limit on $\trh \gtrsim4.3\mev$ which implies an upper limit on $\mmax$ as $\mmax\leq 10.5\ms$ for our benchmark point.}\label{fig:gamma_mmax}
	\end{center}
\end{figure}

This can be seen in figure\,\ref{fig:gamma_mmax} where $\trh$ is plotted as a function of $\Gamma$ in red. This lower limit on $\Gamma$ has crucial implication for $\mmax$ as can be seen from Eq.(\ref{eq:mmax}). The mass $\mmax$ is inversely proportional to $\Gamma$, and this puts an upper limit on $\mmax$ for the minimum value of $\Gamma$. For the benchmark values of parameters discussed in the previous paragraph, $\mmax \lesssim 10.5\ms$. This is shown in figure\,\ref{fig:gamma_mmax} where the blue line represents $\mmax$ as a function of $\Gamma$. Any vertical line drawn in the right hand side of the grey dashed line in this figure is consistent with the EMD scenario, however with a higher value of $\trh$ and therefore, lower value of $\mmax$. We choose to work with a value of $\Gamma$ corresponding to the grey dashed line since our goal is to explore the masses of PBH in the LIGO/Virgo observed range.

\section{Particle dark matter production}\label{sec:particle_production}
\ad{The decay of the moduli $\phi$ \emph{reheats} the Universe. We shall assume all particles in the visible and the dark sectors are produced from the decay of $\phi$.This framework is similar to a normal inflationary reheating model. After $\trh$, when the decay products (visible + dark) start to dominate the total energy density, the Universe transitions from being matter-dominated to radiation-dominated. With this general setup, it is possible to write down myriads of dark sector models\,\cite{Hut:1979xw,Nussinov:1985xr,Okun:1982xi,Georgi:1983sy,Hooper:2007qk,Spergel:1999mh,Loeb:2010gj}. However, to keep our discussion as general as possible, we shall not commit to any particular model. Rather, we shall assume that there exists a dark sector that consists of a DM particle $\chi$ of mass $\mchi$, and one or more lighter states. We characterize the interaction of $\chi$ with other particles using two parameters: the decay branching ratio $b$ of the moduli field to $\chi$, and its thermally-averaged annihilation cross section $\sigmav$. We shall treat $b$ and $\sigmav$ as free parameters in this work. We choose to work with a secluded (or very weakly coupled) dark sector scenario to evade the gamma-ray emission constraints from particle DM accreted around PBHs\,\cite{Carr:2020mqm}.

The decay of $\phi$ to its daughter particles is a continuous process, and during decay, the instantaneous temperature of the radiation bath can be much larger compared to the thermalised reheating temperature. In this case, the dark matter abundance can be produced either through freeze-out or freeze-in mechanism in a matter dominated Universe. In the former case, at early times, the DM particles undergo scattering and are thermalized quickly. Later, a stable abundance of $\chi$ is produced after its annihilation rate $\sigmav$ freezes-out. For the case of freeze-in $\sigmav$ is much smaller compared to the freeze-out case so that the dark matter particle never reaches thermal equilibrium. In this case, the dark matter particle abundance freeze in from the annihilation of other dark sector particles. We shall see later that both scenario produce similar DM relic abundance in the viable region of the parameter space. Moreover, at the end of EMD epoch, dark matter particles can be produced from the direct decay of the $\phi$ particle. 


The decay width of $\phi$ is given in Eq.(\ref{eq;decay_width}).
The abundance of $\chi$ produced from decay of $\phi$ at the end of EMD is given by\,\cite{Gelmini:2006pw,Allahverdi:2018iod}
\beq
\Omega_\mathrm{decay} = \frac{2\pi sb\trh\mchi}{4m_\phi\rho_\mathrm{cr}}\,.
\eeq
Here, $s$ is the entropy and $\rho_\mathrm{cr}$ is the critical density of the Universe. We represent this as a fraction $\fdec$ of the total DM density $\Omega_c$ as
\begin{equation}\label{eq:decay}
\fdec \simeq 0.28 \left(\frac{\mchi}{10\gev}\right) \left(\frac{b}{10^{-4}}\right) \left(\frac{m_\phi}{100\tev}\right)^{1/2}\,.
\end{equation}
Therefore, we see that an $\mathcal{O}(1)$ fraction of $\chi$ population is produced from the moduli decay itself even for a very small branching ratio. 

On the other hand, depending on the value of $\langle \sigma v\rangle$, DM production during EMD epoch can proceed via the following two mechanisms:
\begin{itemize}
	\item \emph{Thermal production:} In this case, $\chi$ interacts with other dark sector or visible sector particles, and quickly thermalizes. A stable population of $\chi$ is finally formed after its annihilation process freezes-out. 
	DM freeze-out during matter domination has been studied before in Refs.\,\cite{Giudice:2000ex,Erickcek:2015jza}. The resulting DM abundance is given by
	\begin{equation}\label{eq:freeze-out}
	\Omega_\mathrm{ann} h^2 = 1.6\times 10^{-4} \dfrac{\sqrt{g_*(\trh)}}{g_*(T_\mathrm{fo})} \left(\dfrac{\mchi/T_\mathrm{fo}}{15}\right)^4 \left(\dfrac{150}{\mchi/\trh}\right)^3 \left(\dfrac{3\times 10^{-26}\cms}{\sigmav}\right)\,.
	\end{equation}
	Here $T_\mathrm{fo}$ is the freeze-out temperature, and $\sigmav$ is the thermally-averaged annihilation cross section. We shall also assume that DM particles are nonrelativistic during freeze-out. From Eq.(\ref{eq:freeze-out}), it is evident that the DM abundance in this case is $\sim 1000$ times smaller than in a radiation dominated Universe.
	
	\item \emph{Non-thermal production:} The $\chi$ particles cannot thermalize if their interactions are very weak. In this case, they are produced from scattering of other particles where the DM matter abundance freezes in non-thermally. This happens when the annihilation cross section $\sigmav$ is much smaller than the previous scenario. The DM abundance is given by\,\cite{Erickcek:2015jza}
	\begin{equation}
	\Omega_\mathrm{ann} h^2 = 0.062\times\dfrac{g_*^{3/2}(\trh)}{g_*^3(\mchi/4)} \left(\dfrac{150}{\mchi/\trh}\right)^5 \left(\dfrac{\trh}{5\gev}\right)^2 \left(\dfrac{\sigmav}{10^{-36}\cms}\right)\,.
	\end{equation}
\end{itemize}
For annihilation cross sections which are in between the two regions, the DM production process would be a combination of these. As a result, one needs to numerically solve the Boltzmann equations for different species. We shall only consider a thermal cross section $\sigmav=3\times 10^{-26}\cms$, which falls in the freeze-out scenario. Similar to decay, we define $\fann \equiv \Omega_\mathrm{ann}/\Omega_c$ as the fraction of total DM abundance that is produced through thermal freeze-out.

These dark sector scenarios are quite general and there are many realizations possible with these characteristics. Therefore, as mentioned before, we shall not go into the details of any particular dark sector model. In passing, we note that the coupling with the moduli $\phi$ itself provides a portal between the visible and the dark sectors. However, a decay width $\Gamma \sim 10^{-23}\gev$ implies that the coupling between $\chi$ and $\phi$ has to be very small. Hence, one needs to consider additional dark sector particles to yield the right DM abundance. These additional interactions introduces the possibility of self-interactions among DM particles\,\cite{Spergel:1999mh,Buckley:2009in,Tulin:2013teo}. There exist strong bounds on DM self-interaction cross section from the observation of dwarf galaxy density profile and galaxy cluster mergers\,\cite{Zavala:2012us,Elbert:2014bma,Kaplinghat:2015aga,Markevitch:2003at}. However, for the coupling and DM mass that yield $\Omega_\chi h^2\simeq 0.12$, the self-scattering cross section is typically below $0.01\,\mathrm{cm^2g^{-1}}$ even when the particles are nonrelativistic\,\cite{Tulin:2013teo,Das:2017fyl}.

\section{Analysis and Results}\label{sec:analysis}
From the discussions in the previous two sections, we see that there are two components of DM in the Universe, PBH and $\chi$ particles, with their fractional abundances $\fpbh$ and $f_\chi$, respectively. The $f_\chi$ again can have two different contributions, $\fdec$ from direct decay of $\phi$, and $\fann$ from annihilation processes. To not overproduce DM, we must ensure
\begin{figure}[!t]
	\begin{center}
		\includegraphics[width=0.48\textwidth]{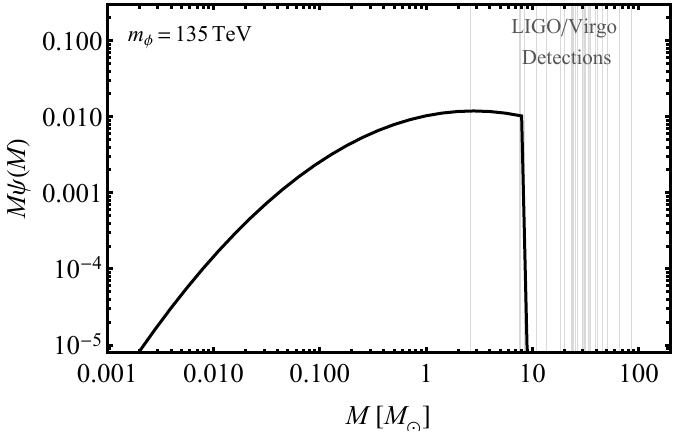}
		\hfill
		\includegraphics[width=0.48\textwidth]{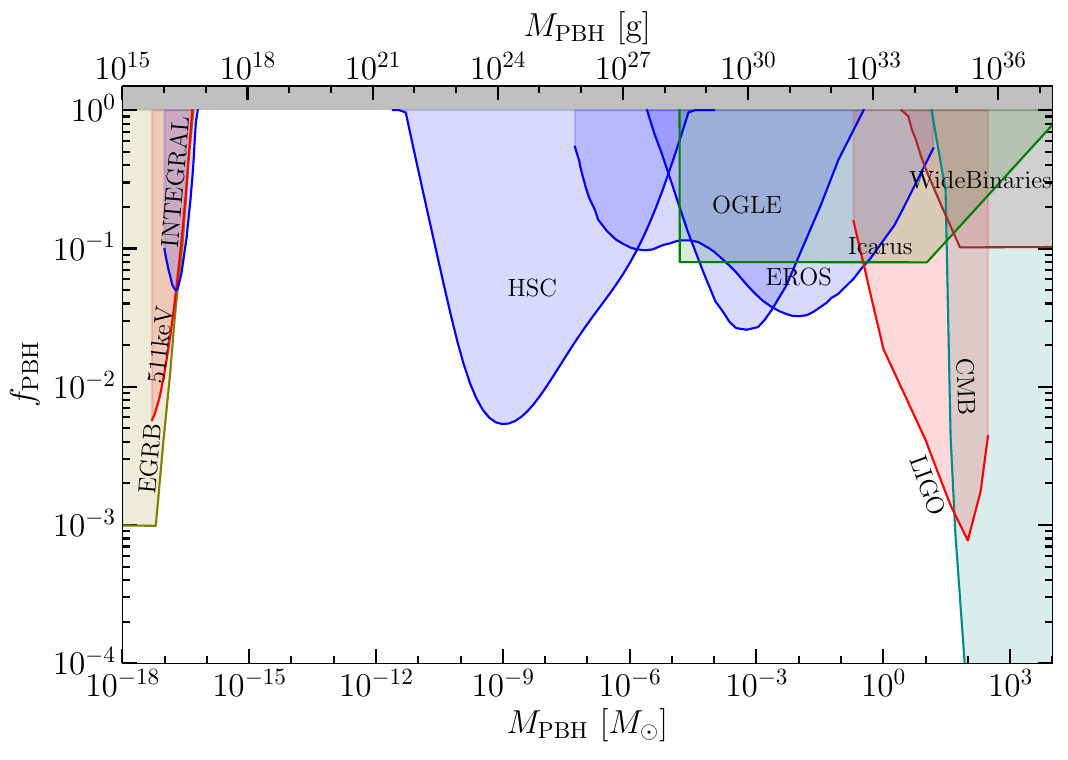}
		\caption{(Left) The PBH mass function for an EMD epoch due to moduli domination are shown for moduli mass $\mphi=135\tev$. The grey vertical lines signify the binary black hole merger events observed by LIGO/Virgo gravitational wave detector. (Right) The excluded regions of the PBH parameter space by various observations and experiments for a monochromatic mass function\,\cite{bradley_j_kavanagh_2019_3538999}. }\label{fig;psiM}
	\end{center}
\end{figure}
\begin{equation}
	\fdm = \fpbh+f_\chi = \fpbh+\fdec+\fann \leq 1
	\label{upconst1}
\end{equation}
There are seven important parameters in our theory, namely, $\mphi,\,\phi_0, A_p, \sigma_p, k_p,\,b,\,\sv$. However, we shall fix $\phi_0=0.05 M_\mathrm{Pl}$ inspired by the conclusions of Ref.\cite{Allahverdi:2018iod}, and vary other parameters. We also fix $k_p=10^6\Mpcinv$ for the reason explained in section\,\ref{sec:PPS}. 
}


The moduli mass $\mphi$ is a crucial parameter as it determines the reheat temperature $\trh$ which denotes the end of the EMD epoch. According to our discussion in section\,\ref{sec:moduli}, saturating the BBN lower bound on $\trh$ yields $\mphi=135\tev$. On the other hand, as we have seen in section\,\ref{sec:PPS}, the largest mass $\mmax$ of the PBHs formed during the EMD era decreases with larger $\Gamma$. The BBN bound therefore implies $\mmax \lesssim 10.5\ms$. From figure\,\ref{fig;psiM}, we see that this is value of $\mmax$ falls near the left end of the BH masses that are observed by the LIGO/Virgo gravitational wave detectors. This value of $\mmax$ is maximal in the sense that, it cannot be increased any further by changing other parameters in $P_\zeta(k)$ as in Eq.(\ref{pow3}). The PBHs of mass above $10.5\ms$, that are observed by LIGO/Virgo, were produced in RD era, and hence their abundance is much smaller relative to the EMD epoch. This can be seen clearly in the left panel of figure\,\ref{fig;psiM}.

\begin{figure}[!t]
	\begin{center}
		\includegraphics[width=0.47\textwidth]{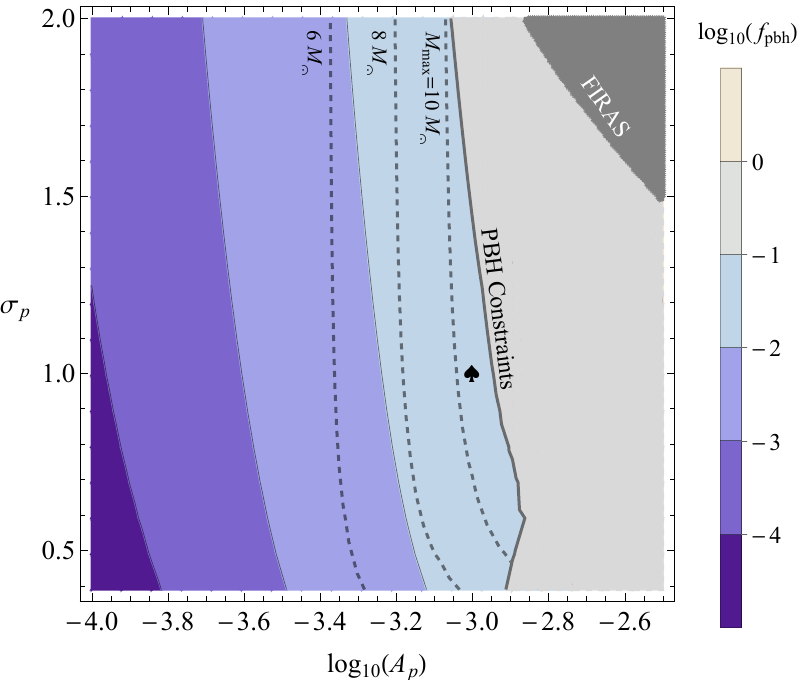}
		~~\quad\includegraphics[width=0.47\textwidth]{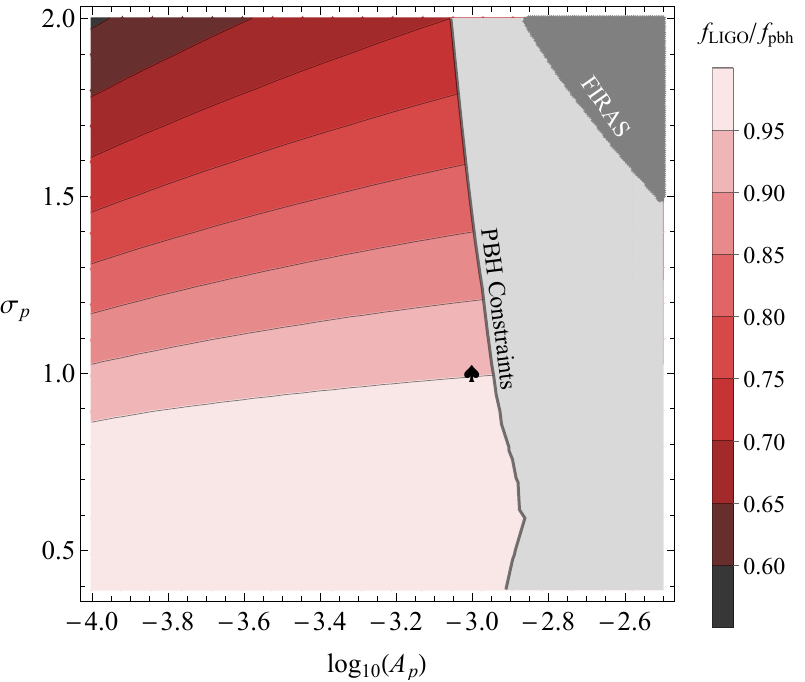}
		\caption{(Left) The variation of $\fpbh$ in the parameter space of $A_p$ and $\sigma_p$ for $\mphi=135\tev$. The light grey-shaded region is excluded by various PBH observation experiments. The dark grey region is ruled out by FIRAS experiment data. The black, dashed lines show the contours of $\mmax=6, 8,$ and $10\ms$ in the same plane. (Right) A contourplot of the ratio $\fligo/\fpbh$, which represents the fraction of the all black holes in the LIGO/Virgo region (see Eq.(\ref{eq:fligo})), in the $A_p-\sigma_p$ plane. Our benchmark point for this work, i.e. $A_p=10^{-3}, \sigma_p=1$, is marked with a spade in both plots.}\label{fig:Ap_sigmap}
	\end{center}
\end{figure}

The PBH mass fraction scales as $\beta(M) \sim \sigma(M)^5$ for in matter domination which gives an enhanced contribution in EMD epoch as compared to RD mass function, which can also be evident from Eq.~\eqref{psiMall}. In EMD epoch, PBHs are formed within a mass range of $\mmin < M < \mmax$. From figure\,~\ref{fig;psiM}, we see that the PBH mass function $\psi(M)$ does not vary by many orders of magnitude within this range\footnote{We did not consider spin of PBH in our calculations, which can have a nonzero value for MD formation in general. Introducing spin modifies $\beta (M)$ for MD epoch~\cite{Harada:2017fjm} for $\sigma (M) < 0.005$, which corresponds to PBH mass $M\lesssim 10^{-7}\ms$ in our case, which contributes negligibly to the total PBH abundance.}. Hence, the most stringent constraints for such a mass function are the microlensing constraint by the Subaru HSC telescope, and the LIGO/Virgo merger event rate. They limit the maximum value of the mass function below $\psi(M) \lesssim 10^{-3}$. The main two parameters that control the shape of $\psi(M)$ are the amplitude $A_p$ and the width $\sigma_p$ of the Gaussian peak in Eq.(\ref{pow3}). The mass scale $\mmin$ representing the lower mass end of the EMD epoch is mainly controlled by initial field value $\phi_0$ of the moduli, and its mass $\mphi$. We fix $\phi_0=0.05\,M_\mathrm{Pl}$ throughout this work. Our final results are not very sensitive to this choice as the mass function decreases towards the smaller $M$ with a subdominant contribution to the integral in $\fpbh$.

In the left panel of figure\,\ref{fig:Ap_sigmap}, we show the PBH fraction $\fpbh$ in the parameters space of $A_p$ and $\sigma_p$. We see that $\fpbh$ increases very sharply with $A_p$ as it changes the height of the Gaussian profile the primordial power spectrum. On the other hand, $\fpbh$ has a relatively weak dependence on $\sigma_p$. The reason is that $\fpbh$, calculated from $\psi (M)$ in EMD, is dominated by the part of $\psi(M)$ close to $\mmax$, and the value of $\mmax$ is anchored by the scale $k_\mathrm{max}$ close to the peak of the Gaussian in $P_\zeta(k)$ by our choice of $\mphi$ (See the left panel of figure\,\ref{fig:primordial_spectrum} and the discussion in section\,\ref{sec:PPS}). Therefore changing $\sigma _p$ does not drastically affect the PBH abundance. 

The light grey-shaded region in figure\,\ref{fig:Ap_sigmap} represents the parameter space ruled out by different observations as discussed in the introduction. 
To apply the constraints, that are conventionally derived for monochromatic mass function, to our extended mass function we use the prescription outlined in Ref.\cite{Carr:2017jsz}. Specifically, we use
\begin{equation}
\int dM~\frac{\psi(M)}{f_\mathrm{max}(M)} \leq 1\,,
\end{equation}
where $f_\mathrm{max}(M)$ is the upper limit on $\fpbh$ from individual experiments. We see that at our benchmark point (marked with a spade), the PBHs form about 4\% of the total DM abundance. The dark grey-shaded region in Figure~\ref{fig:Ap_sigmap} is not allowed from the FIRAS constraints, which are introduced from the bounds on $P_{\zeta}(k)$ itself and independent of the PBH abundance.
The abundance of PBH in the mass range relevant to the LIGO/Virgo detectors could be calculated in the following way:
\begin{equation}\label{eq:fligo}
f_{\rm LIGO} = \int _{0.1\ms}^{300\ms} \psi (M) dM. 
\end{equation}
In the right panel of figure\,\ref{fig:Ap_sigmap}, we show the ratio $\fligo/\fpbh$, which is the fraction of the total PBH population that falls in the LIGO/Virgo mass range, in the same $A_p$-$\sigma_p$ plane. We see that this fraction goes down as $\sigma_p$ is increased. For our chosen benchmark point, about 95\% of total PBH abundance comes from this range. This is due to the fact that the mass function rises steeply towards larger mass end which dominates the integral in Eq.(\ref{eq:fligo}).

\begin{figure}[!t]
	\begin{center}
		\includegraphics[width=0.47\textwidth]{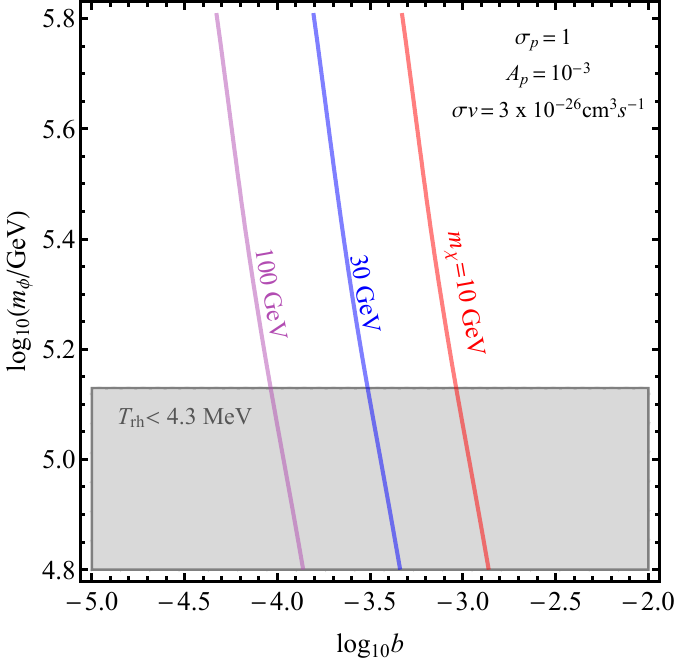}
		\caption{The variation of $\fdm \equiv \fpbh+f_\chi$ in the $b-m_\phi$ parameter plane for three different DM masses $\mchi=10, 30,$ and $100\gev$. The grey-shaded region is excluded by BBN as $\trh < 4.3\mev$. As $\fdm$ is dominated by $f_\chi$, we see that $\fdm$ scales linearly with $b$.}\label{fig:b_mphi}
	\end{center}
\end{figure}

The rest of the dark sector has to be populated by particle DM $\chi$. We have seen before that the branching fraction $b$ of $\phi$, and the annihilation cross section $\sigmav$ determine the present day abundance of $\chi$. In figure\,\ref{fig:b_mphi}, we show the variation of $\fdm\equiv \fpbh+f_\chi$ in the $b$-$m_\phi$ plane. The solid lines corresponds to the contours of $\fdm=1$ for three different DM masses. Values of $\mphi$ smaller than $135\tev$ are excluded as they predict $\trh<4.3\mev$ as discussed earlier. This region is shaded in grey. The upper edge of this grey-shaded region corresponds to our benchmark point where $f_\chi$ dominates over $\fpbh$. This explains the linear dependence of $\fdm$ on $b$ as $\fdec\sim b$. Larger value of $\mchi$ implies smaller $b$ because $\fdec$ is proportional to the product of $b$ and $\mchi$ as can be seen from Eq.(\ref{eq:decay}). 

Lastly, we would like to emphasize two points here. By saturating the BBN bound on low reheating temperature yields $\mphi=135\tev$, which in turn implies $\mmax= 10.5\ms$ as a robust prediction for a fixed value of $k_p=10^6 \Mpcinv$ in such early moduli-domination scenarios. Even for other values of $k_p$, $\mmax$ only decreases which can be seen from the right panel of figure\,\ref{fig:primordial_spectrum}. From figure\,\ref{fig:primordial_spectrum}, \ref{fig;psiM}, and \ref{fig:Ap_sigmap}, we see that the observational bounds on PBHs  are more stringent than the constraints on the primordial spectrum. The FIRAS data excludes only a portion of the parameter space which is already excluded by other observational constraints on PBH abundance.

The analysis in this section and the previous ones emphasise that the 5 parameters that we have not fixed, i.e.,  $m_{\phi}$, $A_p$, $\sigma _p$, $b$ and $\langle \sigma v\rangle$, cannot be varied absolutely freely. Our goal to obtain LIGO mass PBHs fixes the value of $m_{\phi}\simeq 135$ TeV. The inflationary parameters $A_p$ and $\sigma _p$ are to be varied with caution when we aim to reach the maximum possible power around $k_{\rm max}$ due to various experimental bounds on the primordial power spectrum as well as on the PBH abundance (see Fig.~\ref{fig:primordial_spectrum}, right panel of Fig.~\ref{fig;psiM} and Fig.~\ref{fig:Ap_sigmap}). The analysis for PBH production involves only these parameters, whereas the other two parameters $b$ and $\langle \sigma v\rangle$ influence the particle DM content and the dependence can be seen in Fig.~\ref{fig:b_mphi}.

Before concluding, we would like to mention about a recent work that discusses about constraining similar multi-component dark sector scenarios where PBH and particle DM coexist\,\cite{Carr:2020mqm}. There a strong constraint on $\fpbh$ was derived from nonobservation of gamma ray emission from the annihilations of DM particles accreted around the PBHs. However, this is not applicable to a reasonably broad class of DM models, for example, asymmetric dark matter models, secluded dark sector models where DM particles annihilate into other lighter particles etc. Hence, we did not consider this constraint in this work.



\section{Conclusions and Outlooks}\label{sec:conclusions}
In this work, we have examined PBH production in an EMD epoch induced by a moduli field, and a two-component dark sector scenario consisting of PBHs and particle DM. The EMD epoch commences when a moduli field $\phi$ starts oscillating about its potential minimum and comes to dominate the energy density of the Universe. It ends after $\phi$ decays and reheats the Universe by populating the visible and the dark sectors. During the EMD epoch, the radiation pressure vanishes enhancing the PBH formation relative to a radiation-dominated Universe. The big bang nucleosynthesis limits the reheat temperature from below meaning that the EMD epoch must have ended before $T=4.3\mev$. Interestingly, the largest horizon mass scale that became nonlinear at the end of this epoch, falls at the lower end of the black hole masses detected by the LIGO/Virgo collaboration through gravitational wave search. 

We computed the resulting PBH mass function assuming a Gaussian peak in the primordial power spectrum centered at the pivot scale $k_p=10^6\Mpcinv$. We found that the present observational constraints on PBH population excludes primordial power spectrum amplitude $A_p \gtrsim 10^{-3}$. We found that, without violating any of these constraints, we can only produce PBHs of mass $M\lesssim 10.5\ms$. However, because most of the LIGO/Virgo detections are above this mass, this population of black holes cannot possibly explain the observed merger events. This result is robust against any simple variation of other parameters of the theory. 
We note that at this benchmark point of the parameter space, the PBHs form only about 4.3\% of the DM population, and about 95\% of these black holes are above $0.1\ms$.

The rest of the DM population consists of particle DM that were produced from the decay of the moduli during reheating. These DM particles could be a part of a larger dark sector comprising of other lighter particles which interact with the DM. We discussed about two extreme scenarios where the DM particles undergo thermal freeze-out during EMD or they produced through freeze-in from the scattering of other dark sector particles. These two scenarios are realized depending on the interaction strength of the DM particles with particles. We did not resort to any particular particle physics model for the dark sector as simple scenarios such as this one can be realized in many theories. The advantage of considering secluded dark sector models is that the annihilation of the accreted DM particles around the PBHs would not lead to any visible signal in the form of gamma rays or other particles.

It is interesting to note the sensitivity of our analysis to the choice of the primordial power spectrum. The $P_{\zeta}(k)$ that we consider here is parametrised with the amplitude $A_p$ and width $\sigma _p$ of the small-scale Gaussian profile centred at $k_p$. As discussed in section~\ref{sec:PPS}, these parameters are kept at their optimal values obeying direct observational constraints on $P_{\zeta}(k)$ and constraints on the PBH abundance, pushing both $f_{\rm PBH}$ and $\mmax$ to the largest possible value. If one chooses $P_{\zeta}(k)$ involving a small-scale profile which is asymmetric around the peak scale (such as, lognormal power spectrum with an exponential cutoff~\cite{Vaskonen:2020lbd} or broken power law power spectrum~\cite{Deng:2016vzb,Liu:2019lul,Deng:2017uwc,Cotner:2016cvr,Cotner:2017tir}), then we expect $\mmax$ to have a maximum value very close to our result $10.5\ms$ since $k_{\rm max}$ is very close to $k_p$. On the other hand, $f_{\rm PBH}$ may become larger (smaller) with a less (more) steep slope of the large-$k$ arm of $P_{\zeta}(k)$. This slope could be contrained by the the PTA observations (see left panel of figure~\ref{fig:primordial_spectrum}). Constraints on PBH abundance on lower masses also limits this slope, similar to what has been discussed in section~\ref{sec:analysis} for our figure~\ref{fig:Ap_sigmap}. However, we still expect the resulting $f_{\rm PBH}$ to be of the same order as in our analysis, since lower PBH masses contribute nominally to the integral in $f_{\rm PBH}$.

Several future experiments can potentially constrain this model. The PIXIE experiment is planned to measure the CMB spectral distortions with much higher precision\,\cite{2011JCAP...07..025K}. Its data would exclude a large region of the primordial power spectrum parameter space and improve over the FIRAS constraint by about four orders of magnitude. Future data from SKA will improve the PTA bound by more than two orders of magnitude, and could potentially rule out such small scale Gaussian features in the primordial power spectrum. In the direction of PBH observation, lensing of fast radio bursts could be used to constrain $10^{-4}\ms\lesssim M \lesssim 0.1\ms$ mass objects\,\cite{Katz:2019qug}. Future planned space-based gravitational wave detectors LISA, DECIGO will be sensitive to the level of $\fpbh\simeq 10^{-3}$ for smaller mass black holes in the range $10^{-6}\ms\lesssim M \lesssim 1\ms$\,\cite{Guo:2017njn,Wang:2019kzb}. Survey telescopes, like Gaia and Theia in future, could look for PBHs through their effect on the motions of other nearby stars\,\cite{Dominik_2000,VanTilburg:2018ykj}. In summary, many future observations are planned that will constrain the theory parameter space of such models in the coming decade.

Early matter dominated epoch can arise in scenarios other than moduli domination~\cite{Dror:2016rxc,Berlin:2016gtr,Dror:2017gjq}, each of which offers individual frameworks and set of parameters. Moreover, it will be interesting to perform similar analysis with other forms of the primordial power spectrum, preferably starting from a particular model of inflation. We intend to approach these problems in future projects.

To conclude, in this paper, we have considered a early matter dominated universe induced by a moduli field and explored the possibility of generating solar mass range PBHs with the motivation to explaiin observed events in LIGO/Virgo detectors. We found that for a Gaussian primordial fluctuations at small scales with favorable parameters, the maximum PBH mass is $\sim 10\ms$, and PBH contribution to the total DM is only a few percent. 

\section*{Acknowledgment}
SB is supported by the institute postdoctoral fellowship at Physical Research Laboratory, India and by the National Postdoctoral Fellowship of the Science and Engineering Research Board (SERB), Department of Science and Technology (DST), Government of India (GOI) at different stages of this work. AD is supported by the U.S. Department of Energy under contract number DE-AC02-76SF00515. KD is supported in part by the grant MTR/2019/000395 and Indo-Russian project grant DST/INT/RUS/RSF/P-21, both funded by the DST, Govt of India. 
AD acknowledges the hospitality of the Theory Division of Saha Institute of Nuclear Physics during the initial part of this work. The authors would like to thank Avik Banerjee, Ranjan Laha, Biswarup Mukhopadhyaya and Luca Visinelli for discussions related to the project and help improving the manuscript. The authors would like to thank the organizers of the  Less Travelled Path of Dark Matter: Axions and Primordial Black Holes (code: ICTS/LTPDM2020/11) at the International Centre for Theoretical Sciences (ICTS) for many stimulating discussions. 

\bibliographystyle{jhep}
\bibliography{emd}
\end{document}